\documentclass{cmmse2014}
\usepackage{amssymb}
\usepackage[pdftex]{graphicx}
\usepackage{epstopdf}
\begin{document}



\newcommand{\Eqref}[1]{(\ref{#1})}




\title{On wind Turbine failure detection from measurements of phase currents: a permutation entropy approach}



\author{Sumit Kumar Ram}{skr12ms039@iiserkol.ac.in}{{1,2}}

\author{Geir Kulia}{kulia@stud.ntnu.no}{3}

\author{Marta Molinas}{marta.molinas@ntnu.no}{1}


\affiliation{1}{Department of Engineering Cybernetics}
  {Norwegian University of Science and Technology}

\affiliation{2}{Department of Physical Sciences}{Indian Institute of Science Education and Research, Kolkata}

\affiliation{3}{Department of Electronics and Telecommunications}
  {Norwegian University of Science and Technology}



\begin{abstract}

This article presents the applicability of Permutation Entropy based complexity measure of a time series for detection of fault in wind turbines. A set of electrical data from one faulty and one healthy wind turbine were analysed using traditional Fast Fourier analysis  in addition to Permutation Entropy analysis to compare the complexity index of phase currents of the two turbines over time. The 4 seconds length data set did not reveal any low frequency in the spectra of currents, neither did they show any meaningful differences of spectrum between the two turbine currents. Permutation Entropy analysis of the current waveforms of same phases for the two turbines are found to have different complexity values over time, one of them being clearly higher than the other. The work of Yan et. al. in \cite{Yan} has found that higher entropy values related to the presence of failure in rotary machines in his study. Following this track, further efforts will be put into relating the entropy difference found in our study to possible presence of failure in one of the wind energy conversion systems. 

\end{abstract}

\section{Introduction}

Wind energy conversion systems is the fastest-growing
source of new electric generation in the world and it is expected
to remain so for some time\cite{Amirat}. In order to be more reliable
and competitive than classical power generation systems and
due to geographical location of wind turbines,
it is important to prevent failure and to reduce maintenance
cost. Hence, it becomes important task to categorize the the failed turbines and take necessary actions to prevent further problems in the due process. Traditionally various analysis techniques such as FFT, Wavelet transformation, time-scale decompositions and AM/FM demodulation, have been used to classify the failed turbines from the healthy ones. However, most of the mentioned analysis techniques uses predefined parameters for the analysis and in some cases fail to detect major failures. Yet, recently developed adaptive data analysis techniques such as HHT\cite{huang}\cite{Amirat} and Permutation entropy\cite{Bandt} shows a promising future in the field of advanced fault detection.\par
 Conceptually simple and easily calculated measure of permutation entropy proposed by Bendt et al.\cite{Bandt} can be effectively used to detect qualitative and quantitative dynamical changes. In other words it serves as a complexity measure for time series data, considering the local topological structure of the time series, it assigns a complexity value to each data point\cite{zanin}. Due to its simplicity and robustness it has gained a remarkable popularity in the field of biomedical sciences\cite{Morabito}\cite{Nicolaou}. However the recent advances in its applicability in fault detection in rotary machines\cite{Yan}, bearing fault diagnosis\cite{Wu} has prompted curiosity in further application of this methodology in advanced fault detection mechanisms. In this article we make a simple demonstration of this powerful technique for characterising the wind turbines based on their complexity value of their current waveforms measurement.

\section{Method}
Sample data of length 4 sec consisting of waveforms of $L_1$, $L_2$, $L_3$ phase currents with respect to the ground was collected from two wind turbines named as turbine $T_{14}$ and turbine $T_3$. Table\ref{table} represents the nomenclature of the current phases for the given data. In the first step of our analysis the FFT for each current waveforms was calculated, compared and analysed to detect if the failure could be due to some problems related to the bearings. In the next step, the data set was analysed using permutation entropy values of the currents to compare the complexity based on the local topological variation of both turbines.   

\begin{table}[h!]
\centering
\caption{Defining $L_1$,$L_2$ and $L_3$}
\label{table}
\begin{tabular}{|l|l|l|l|l}
\hline
Name  & Current(A) & Description \\
\hline
$L_1$ & Phase a    &      $L_1$,$L_2$ and $L_3$                        \\
$L_2$ & Phase b    &    represents the  current waveforms               \\
$L_3$ & Phase c    &    for phase lines a,b and c respectively           \\
\hline
\end{tabular}
\end{table}

\subsection{FFT Analysis}
Fast Fourier Transform for phase currents $L_1$,$L_2$ and $L_3$ was calculated and corresponding six major amplitude and frequency spectrum were noted down. The amplitude and frequency value for each phase was compared for both turbines $T_{14}$ and $T_3$. Furthermore the basic purpose of the analysis was to detect any major low frequency component which are usually the result of bearing failure and whether there is a measurement difference in both turbines' frequency spectra of current waveforms. This difference, if exists could be an indicator of possible failures.  

\subsection{Permutation Entropy Analysis}
The permutation entropy value of $L_1$,$L_2$ and $L_3$ over time with the parameter values: sequence length ($w$)=400,time delay($\tau$)=1 and embedding dimension($m$)=3 was calculated for each phase of the given turbines $T_3$ and $T_{14}$. For more details about the permutation entropy calculation and parameter selection refer to the internal report\cite{sumit} submitted at Department of Engineering Cybernetics by Balchen Scholarchip grantee Sumit Kumar Ram.

\section{Results}

\subsection{FFT Analysis}
\begin{figure}
    \centering
    \includegraphics[scale=0.5]{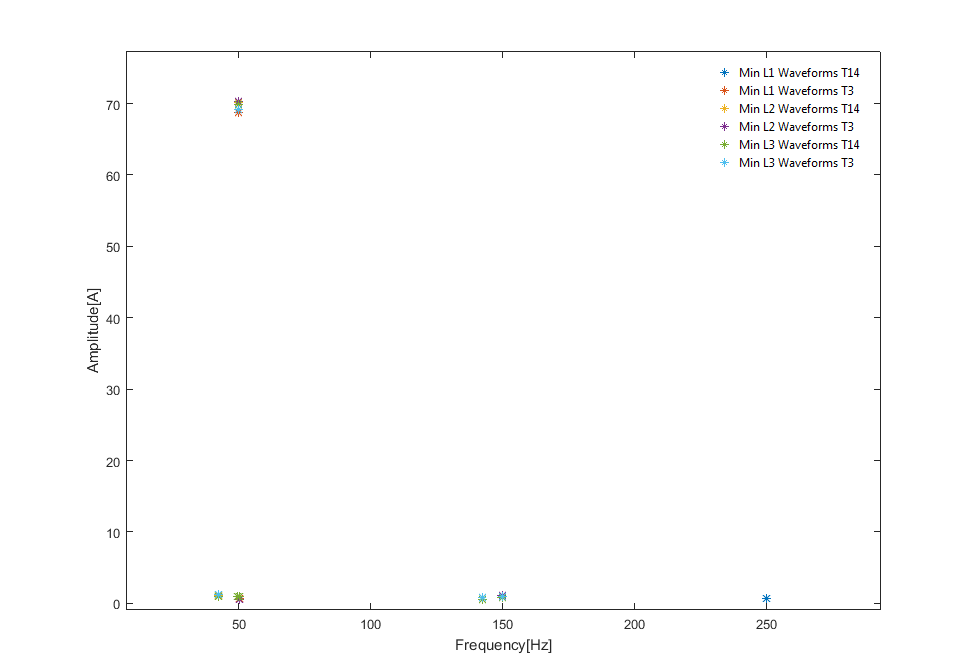}
    \caption{Amplitude vs Frequency plot for the $L_1$,$L_2$,$L_3$ current waveforms of turbine 14 and 3}
    \label{fft_all}
\end{figure}

As it can be observed from fig.\ref{fft_all}, all the phases shows normal behavior in terms of their frequency and amplitude spectra of the currents. The phases do not show any low frequency component other than 42Hz. Due to low amplitude value higher frequencies are very difficult to be detected from the FFT analysis which may have caused due to malfunction in the working condition of the turbine as we concluded but FFT does not provide us useful information regarding this. Next step of of the analysis was carried out using permutation entropy analysis to investigate the minute changes in the time series data which can be compared to conclude the working condition of the turbine.\par 
The given data set is of very short duration, so FFT is unable to detect the low frequency component even if it has a higher amplitude. Hence FFT cannot detect bearing fault from the measured current waveforms at the current point, even if it is present and has a very low frequency value.

\subsection{Permutation Entropy Analysis}
\begin{figure}
\centering
\includegraphics[scale=0.5]{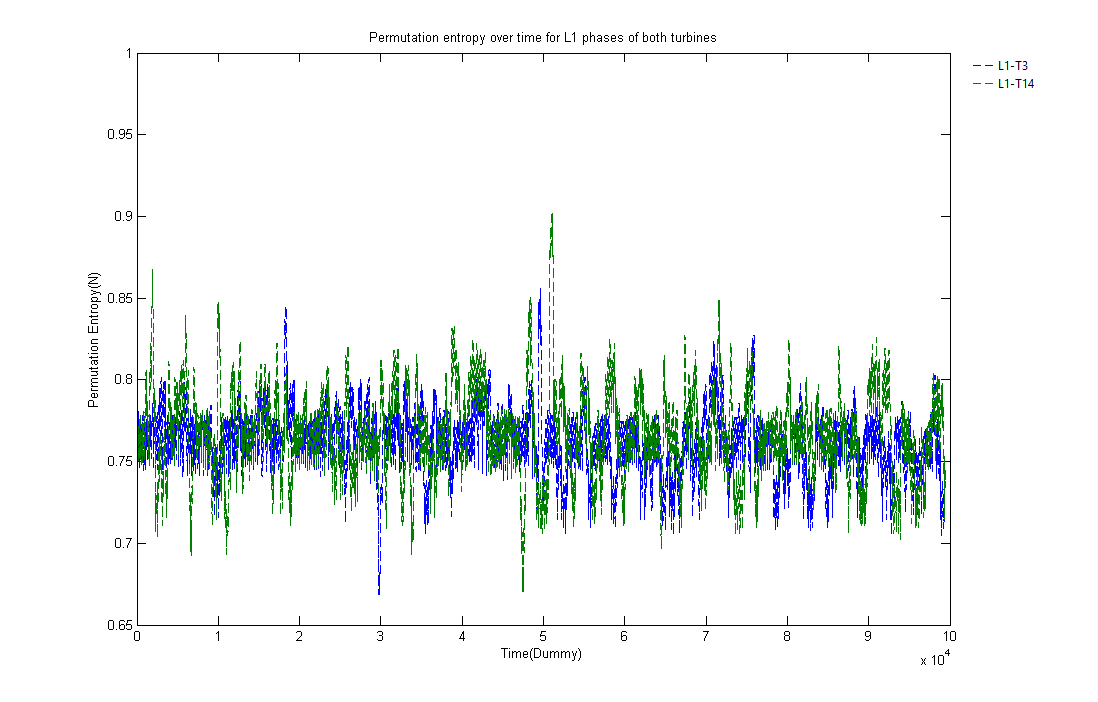}
\caption{Comparison of permutation entropy values for $L_1$ min waveforms of both turbine 3 and 14.}
\label{l1_both}
\end{figure}

\begin{figure}
\centering
\includegraphics[scale=0.5]{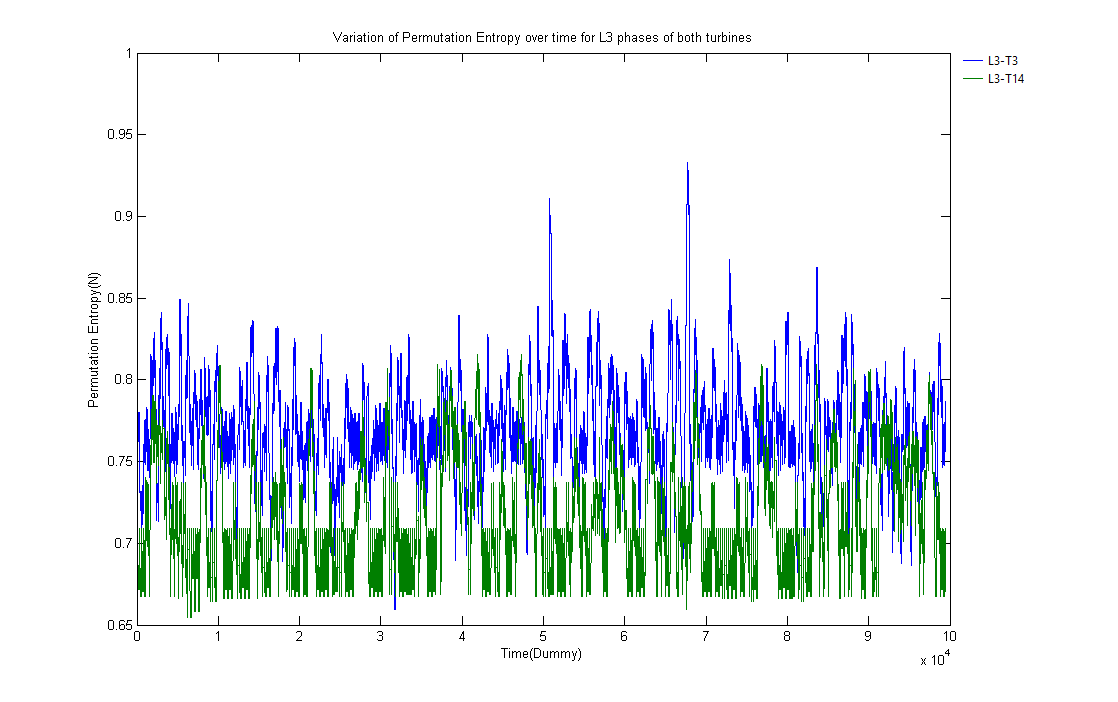}
\caption{Comparison of permutation entropy values for $L_3$ min waveforms of both turbine 3 and 14.}
\label{l3_both}
\end{figure}

\begin{figure}
\centering
\includegraphics[scale=0.5]{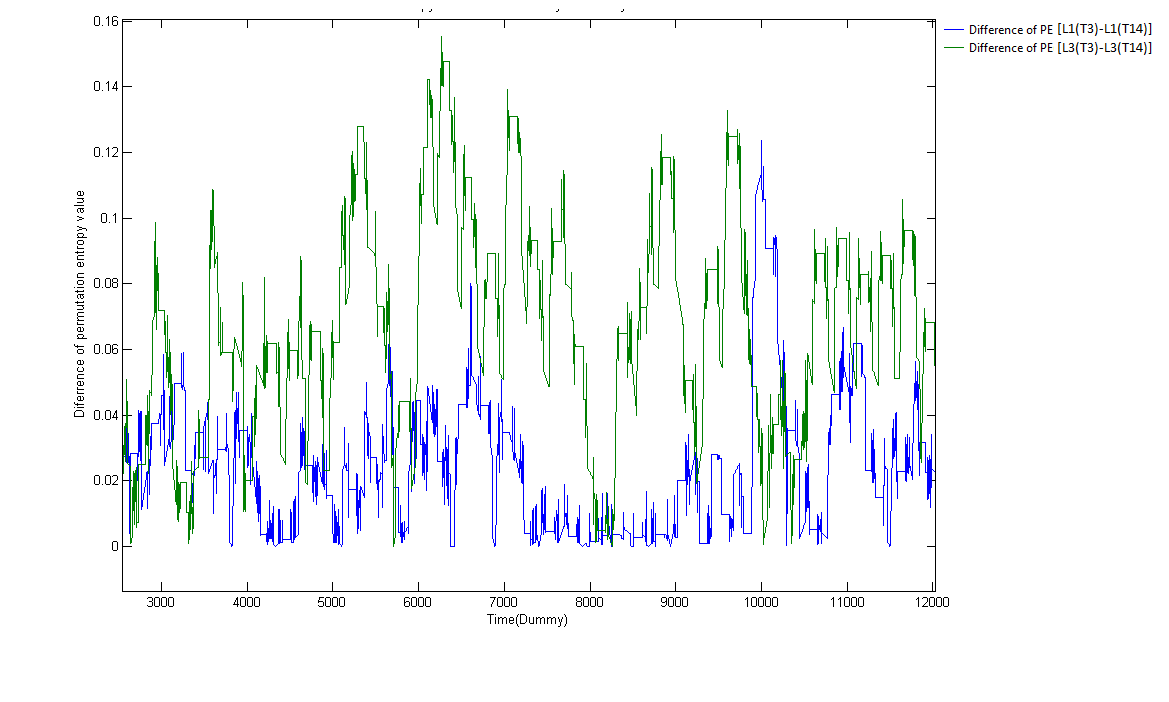}
\caption{Difference of permutation entropy values of $L_1$ and $L_3$ waveforms for turbines 3 and 14.}
\label{diff_both}
\end{figure}

 The permutation entropy value for each current waveforms was calculated for both turbines and was compared for analysis. It was found that for most of the current waveforms of both turbines the permutation entropy values are comparable. \textbf{But, there exists some waveforms for which the permutation entropy value does not match and the waveforms for one turbine has a comparatively higher permutation entropy value with respect to the corresponding waveforms of the other turbine.}  Fig. \ref{l1_both} and Fig. \ref{l3_both} illustrates the above statement taking into account the permutation entropy values of both turbines $T_{14}$ and $T_3$ for $L_1$ and $L_3$ waveforms. It can be seen that for the same waveforms ($L_1$) of both turbines $T_{14}$ and $T_3$, the permutation entropy value can be comparable, meaning, they have almost same average and standard deviation through out the time frame. Where as the permutation entropy value for $L_3$ waveforms of both turbines in fig. \ref{l3_both} has different values for each turbine and are not comparable with each other. Fig. \ref{diff_both} shows the difference of permutation entropy values for both $L_1$ and $L_3$ waveforms from both the turbines. It is clear from the figure that the difference of permutation entropy values for the $L_1$  waveforms has comparatively lower value than $L_3$ waveforms which can be infered as the indication of working condition of the turbines being different.  
 
\section{Discussion}
Permutation entropy serves as a parameter to classify the turbines based on their complexity value of their current waveforms. This can act as a classifier and can be coupled with machine learning methodology along with other statistical analysis to develop an algorithm which can detect the working condition of the turbine and give information about the potential hidden failures. The entropy measurement is robust and is computationally affordable for application in real time.

\section*{Acknowledgments}
The authors would like to acknowledge the companies Kongsberg Maritime and Sintef for providing raw data, as well as the Department of Engineering cybernetics, NTNU for the financial support through Jans Balchen scholarship, without which the research work would not have been possible.


\begin{thebibliography}{1}

\bibitem{Amirat}Amirat, Yassine, Vincent Choqueuse, and Mohamed Benbouzid. "EEMD-based wind turbine bearing failure detection using the generator stator current homopolar component." Mechanical Systems and Signal Processing 41.1 (2013): 667-678.

\bibitem{huang}Huang, Norden E., et al. "The empirical mode decomposition and the Hilbert spectrum for nonlinear and non-stationary time series analysis." Proceedings of the Royal Society of London A: Mathematical, Physical and Engineering Sciences. Vol. 454. No. 1971. The Royal Society, 1998.

\bibitem{sumit}Ram, Sumit K., Kulia, Geir, Molinas, Marta. "Analysis of healthy and failed wind turbine electrical data: apermutation entropy approach." Department of Engineering Cybernetics, Internal publications,  NTNU. Jan 2016.

\bibitem{Bandt}Bandt, Christoph, and Bernd Pompe. "Permutation entropy: a natural complexity measure for time series." Physical review letters 88.17 (2002): 174102.

\bibitem{zanin}Zanin, Massimiliano, et al. "Permutation entropy and its main biomedical and econophysics applications: a review." Entropy 14.8 (2012): 1553-1577.

\bibitem{Morabito}Morabito, Francesco Carlo, et al. "Multivariate multi-scale permutation entropy for complexity analysis of Alzheimer’s disease EEG." Entropy 14.7 (2012): 1186-1202.

\bibitem{Nicolaou}Nicolaou, Nicoletta, and Julius Georgiou. "Detection of epileptic electroencephalogram based on permutation entropy and support vector machines." Expert Systems with Applications 39.1 (2012): 202-209.


\bibitem{Yan}Yan, Ruqiang, Yongbin Liu, and Robert X. Gao. "Permutation entropy: A nonlinear statistical measure for status characterization of rotary machines." Mechanical Systems and Signal Processing 29 (2012): 474-484.

\bibitem{Wu}Wu, Shuen-De, et al. "Bearing fault diagnosis based on multiscale permutation entropy and support vector machine." Entropy 14.8 (2012): 1343-1356.
\end{thebibliography}
\end{document}